# Detection and Imaging of High-Z Materials with a Muon Tomography Station Using GEM Detectors

K. Gnanvo, B. Benson, W. Bittner, F. Costa, L. Grasso, M. Hohlmann, *Member, IEEE,* J.B. Locke, S. Martoiu, H. Muller, M. Staib, A. Tarazona, and J. Toledo

*Abstract*–Muon tomography based on the measurement of multiple scattering of atmospheric cosmic ray muons is a promising technique for detecting and imaging heavily shielded high-Z nuclear materials such as enriched uranium. This technique could complement standard radiation detection portals currently deployed at international borders and ports, which are not very sensitive to heavily shielded nuclear materials. We image small targets in 3D using $2\times2\times2$ mm$^3$ voxels with a minimal muon tomography station prototype that tracks muons with Gas Electron Multiplier (GEM) detectors read out in 2D with *x-y* microstrips of 400 μm pitch. With preliminary electronics, the GEM detectors achieve a spatial resolution of 130 μm in both dimensions. With the next GEM-based prototype station we plan to probe an active volume of ~27 liters. We present first results on reading out all 1536 microstrips of a $30\times30$ cm$^2$ GEM detector for the next muon tomography prototype with final frontend electronics and DAQ system. This constitutes the first full-size implementation of the Scalable Readout System (SRS) recently developed specifically for Micropattern Gas Detectors by the RD51 collaboration. Design of the SRS and first performance results when reading out GEM detectors are presented.

*Keywords*: Muon Tomography; Multiple Scattering; MPGD; GEM Detector; High-Z materials; Scalable Readout System

## I. Introduction

Muon Tomography (MT) exploits the significant multiple Coulomb scattering of atmospheric cosmic ray muons in objects with high atomic number Z for detecting and imaging high-Z nuclear materials in cargo, such as highly enriched uranium. Due to the penetrating nature of multi-GeV cosmic ray muons, this technique is a good candidate for defeating concealment of nuclear contraband with heavy shielding. In general, muon tomography requires tracking the cosmic ray muons before they enter and after they exit the probed volume so that the scattering angles of muons traversing the volume can be measured precisely. The tracking detectors that are being investigated specifically for muon tomography applications are mainly gaseous detectors because these detectors can cover large areas while offering good efficiency and spatial resolution for muon detection. Examples are MT applications of drift tubes [1]-[5], drift chambers [6], [7], and resistive plate chambers [8].

Our effort focuses on using micropattern gas detectors (MPGDs), specifically Gas Electron Multiplier (GEM) detectors [9], as tracking detectors for muon tomography [10], [11] because MPGDs can achieve the highest spatial resolutions (50 to 150 μm) among gas detectors. This approach should allow the construction of quite compact muon tracking stations of only ~10 cm thickness, whereas the other types of gas detectors typically require tracking stations of considerably larger vertical size (~1 m) above and below the probed volume. The challenge in the application of GEM detectors to muon tomography is to develop GEM detectors that can cover the required large areas and electronics that can handle the large number of required readout channels.

Development of large-area MPGDs and of a readout system that can be easily scaled up from a few hundred channels for small test systems to over 100k channels for large systems are two of the main R&D thrusts pursued by the RD51 collaboration [12] at CERN. We are participating in this RD51 effort and are presenting here the first results of a complete readout achieved for a GEM detector (1536 channels) with such a Scalable Readout System (SRS) for MPGDs.

## II. Results from Minimal Muon Tomography Station

We previously reported on design and construction [13] of several Triple-GEM detectors with 30 cm × 30 cm active area for a small prototype of a GEM-based muon tomography station (MTS). First muon tomography results were obtained by reading out minimal areas of 5 cm × 5 cm in the centers of two such detectors placed above the target volume and two below using preliminary electronics. These results focused on 2D reconstruction of small target objects [14]. Here we present results on the measurement of the spatial resolution after aligning the four detectors with muon tracks and on the 3D reconstruction of the targets with that minimal MTS setup.

### A. Detector alignment with muon tracks

Muon tracks recorded with the minimal MTS with a 4 mm thick pressboard target support plate present in the active volume are used to align the detectors. This "empty" scenario is used for alignment because the muon tracks will be minimally affected by the pressboard and can be fit to straight lines to good approximation. Due to electronics limitations, only half of one dimension of each detector is read out at maximum spatial resolution. This region is the "single-strip region," where one readout strip is connected to one readout

---

Manuscript received November 13, 2010. This work was supported in part by the U.S. Department of Homeland Security under Grant No. 2007-DN-077-ER0006-02.

K. Gnanvo, B. Benson, W. Bittner, L. Grasso, M. Hohlmann, J. B. Locke, and M. Staib are with the Department of Physics and Space Sciences, Florida Institute of Technology, Melbourne, FL 32901, USA (telephone: 321-674-7275, e-mail: kgnanvo@fit.edu); F. Costa, S. Martoiu, and H. Muller are with CERN, Geneva, CH 1211, Switzerland; A. Tarazona and J. Toledo are with I3M Institute, Universidad Politécnica de Valencia, Valencia, Spain.

channel. The other region is the "double-strip region," where two ganged strips are read out by one electronics channel.

The relative alignment of the four GEM detectors is performed using only the single-strip regions for $x$ and $y$. Each muon track is fit to straight lines in the $x$-$z$ and in the $y$-$z$ planes, which makes the alignments in $x$ and in $y$ independent. There are 188 usable muon tracks in $x$ and 233 usable muon tracks in $y$. The residuals between every detector hit and the fit line are recorded separately in both dimensions and for each of the four detectors. The means of these distributions are listed in Tab. 1 in the "Initial Mean" column. The values of the means of the residual distributions are subtracted from the corresponding data points. The corrected data are refit, and the corrected residual distributions are found to be centered closer to zero as expected.

The residual distributions corrected for alignment are shown as data points in Fig. 1, and their means are listed in Tab. 1. Continued iterations of this process are not found to bring the centers of the residual distributions closer to zero on average, so the initial means in Tab. 1 are taken as the best estimates of the physical misalignments of the detectors. All subsequent reconstructions of multiple scattering for the real data have these offsets implemented for each detector.

### B. Spatial resolution measurement

The minimal MTS is simulated with the Monte Carlo simulation toolkit GEANT4 [15]. The detector misalignments observed in the real data are implemented in the simulation, and then the simulated data are aligned in the same way as the real data. Only data from the single-strip regions are used to determine the spatial resolution. Muon tracks are simulated for spatial detector resolutions of 100 to 200 μm in 10 μm increments. We compare the corrected residual distributions for each detector in each dimension for simulated and real data. The average differences between the real and simulated means and standard deviations of the corrected residual distributions are recorded and the minimum difference is found. The closest match between simulated and real data is found for a simulated spatial detector resolution of 130 μm. The corrected residual distributions for real and simulated data with 130 μm detector resolution are shown in Fig. 1. We find 130 μm as our best estimate of the spatial resolution achieved by the Triple-GEM detectors with preliminary electronics.

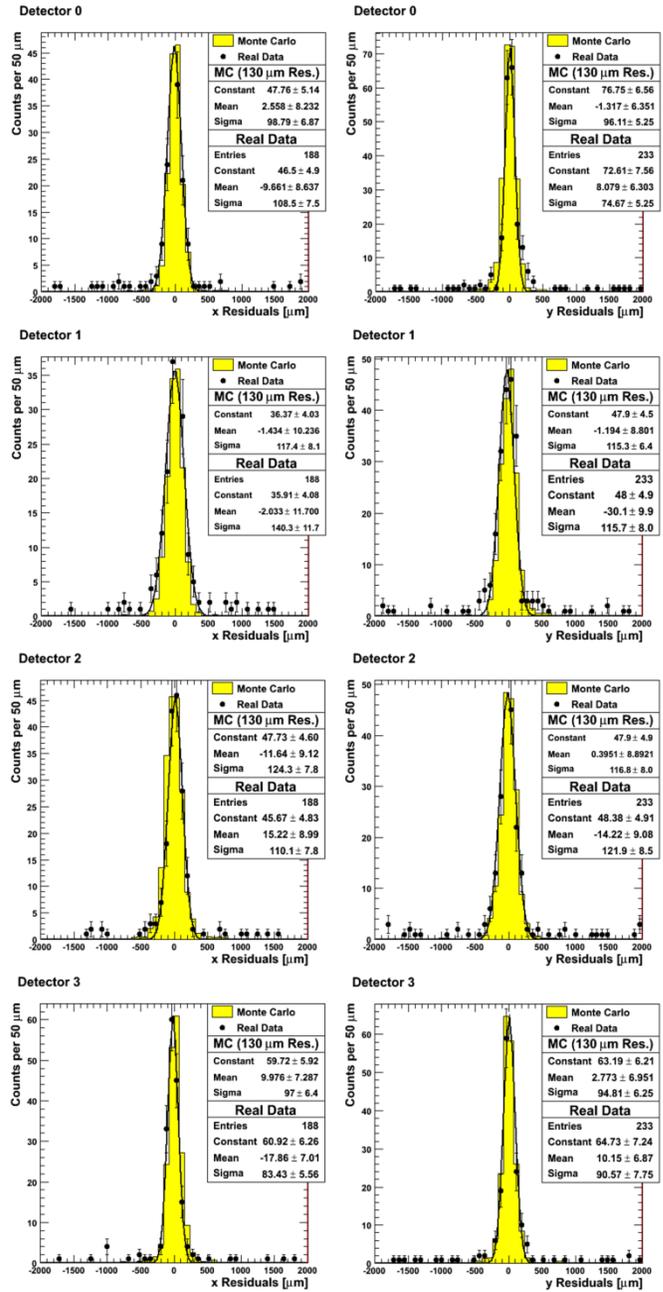

Fig. 1. Residual distributions for data (points) and Monte Carlo simulation for spatial detector resolution of 130 μm (histograms) after detector alignment. MC distributions are normalized to the maximum bin in the corresponding data. Detectors are numbered 0–3 from top to bottom.

TABLE I. MEANS OF INITIAL AND ALIGNED RESIDUAL DISTRIBUTIONS FOR DATA. DETECTORS ARE NUMBERED 0–3 FROM TOP TO BOTTOM.

| Detector [#] | Dimension [$x$ or $y$] | Initial Mean [μm] | Aligned Mean [μm] |
|---|---|---|---|
| 0 | $x$ | 151 | -10 |
| 0 | $y$ | 110 | 8 |
| 1 | $x$ | -156 | -2 |
| 1 | $y$ | -97 | -30 |
| 2 | $x$ | -141 | 15 |
| 2 | $y$ | -134 | -14 |
| 3 | $x$ | 134 | -18 |
| 3 | $y$ | 125 | 10 |

### C. 3D reconstruction and tomographic imaging

A $30 \times 30 \times 30$ mm$^3$ iron cube, a $28 \times 20 \times 30$ mm$^3$ lead block, and a tantalum cylinder of 30 mm diameter and 16 mm height are used as test targets for the minimal MTS [14]. We reconstruct scattering points and angles using the same simple Point-of-Closest-Approach (POCA) algorithm [10], [11] that we previously applied to the minimal MTS data for 2D reconstruction and target imaging [14]. The main difference is that the data are now corrected for detector misalignment using the results described in the previous section.

We have developed stand-alone 3D voxel visualization software "Vx4D" based on OpenGL that can display voxelized data in color. This software displays all voxels at the same size with voxel color and transparency representing the weight of a voxel. We have created this custom graphics software because CERN's standard ROOT [16] data analysis and display package currently cannot handle 3D histograms in OpenGL with such features [17].

By developing a stand-alone software package we are able to add useful features such as the ability to load nominal target volumes into the visualization for comparison with the reconstructed data. We can apply cuts that can be adjusted interactively, e.g. on muon number or scattering angle, so that the results can be better visualized. The software accepts mouse and keyboard input to allow the user to rotate the volume on all 3 axes and to zoom in and out. The user also has the ability to "fly" through the volume.

The software is written in C++, object-oriented, and runs on Linux x86/64. It is multi-threaded and has an interface written in GTK. It provides an extensible algorithmic interface to allow users to implement their own algorithms in the display, e.g. a clustering or threshold algorithm.

3D color images of the three small target objects using ~1000 tracks recorded with the minimal muon tomography station and displayed with Vx4D, i.e. "muon tomograms," are shown in Figs. 2-4 using small 2×2×2 mm$^3$ voxels. A spatial clustering of voxels with large scattering angles is clearly observed in the regions of the actual target locations. The high-Z, high-density Ta target shows the most clustering as expected. For comparison, Fig. 5 shows a high-statistics Monte Carlo simulation of the Ta target displayed in the same way. For this simulated muon tomogram, a cut of $\Theta_{mean} > 1°$ is applied for the sake of clearly displaying the central target region. The observed vertical blurring in the image is caused by the lack of muons probing the targets from the side because the minimal MTS mainly accepts vertical muons due to its detector arrangement and small active areas. The next prototype should overcome this limitation by using side detectors and a much larger angular acceptance.

### III. SCALABLE READOUT SYSTEM FOR MPGDs

The Scalable Readout System (SRS) [18] is an RD51 initiative [12] to establish a "Portable Multichannel Readout System for Micropattern Gas Detectors" including protection of the on-detector electronics against discharges that can occur in gaseous detectors. Proposed in 2009 as a scalable readout architecture with a general-purpose chip frontend, the first complete small SRS prototype has become available in the fall of 2010. It consists of 12 front-end hybrid boards based on an analog amplifier chip, which are connected via HDMI cables to a pair of ADC/Frontend-Concentrator electronics cards as shown in Fig. 6. A Gb ethernet link connects the small-system SRS electronics to a Linux PC running DATE online data acquisition software [19]. The ROOT-based AMORE package [20] is used for online and offline data analysis.

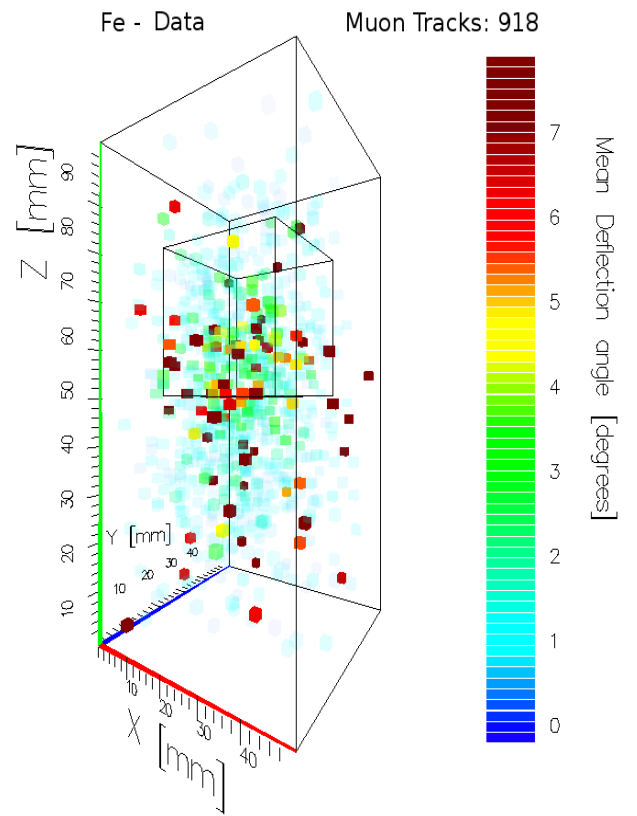

Fig. 2. Muon tomogram of a Fe cube with Vx4D using 918 muon tracks.

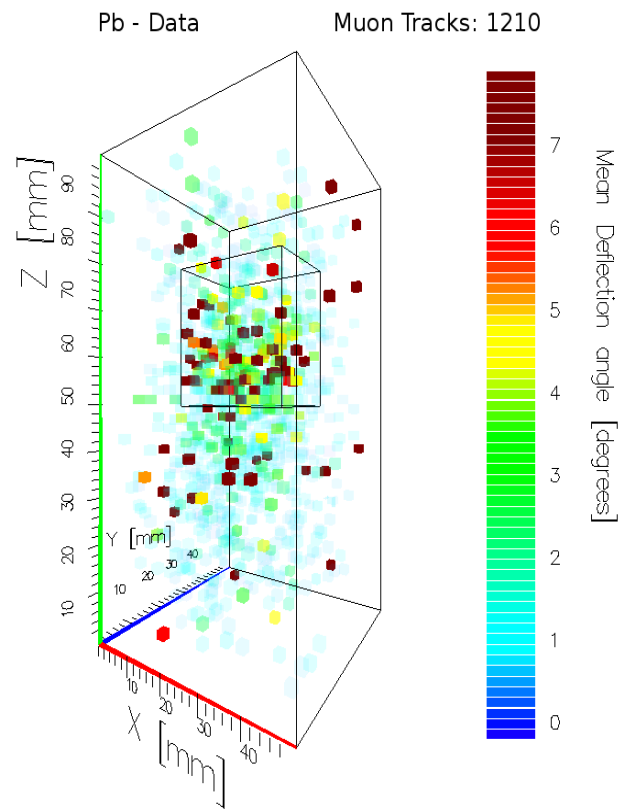

Fig. 3. Muon tomogram of a Pb block with Vx4D using 1210 muon tracks.

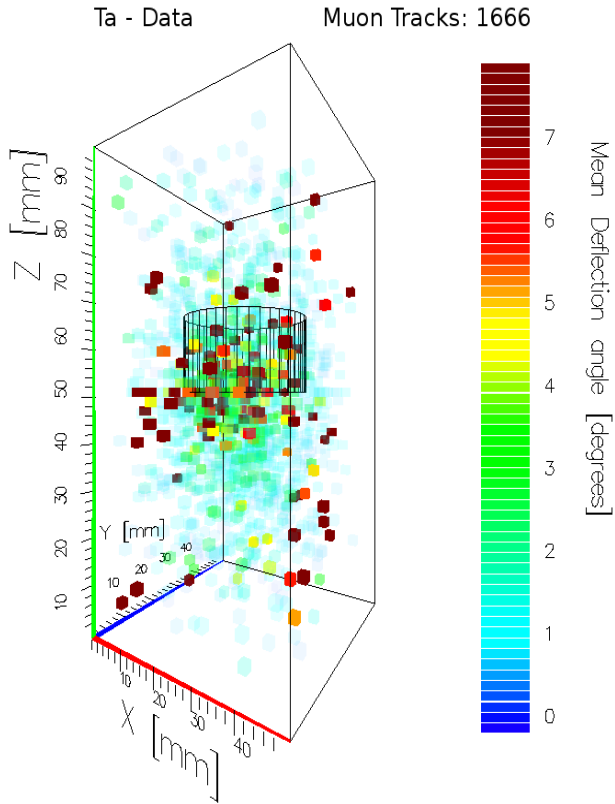

Fig. 4. Muon tomogram of Ta cylinder with Vx4D using 1666 muon tracks.

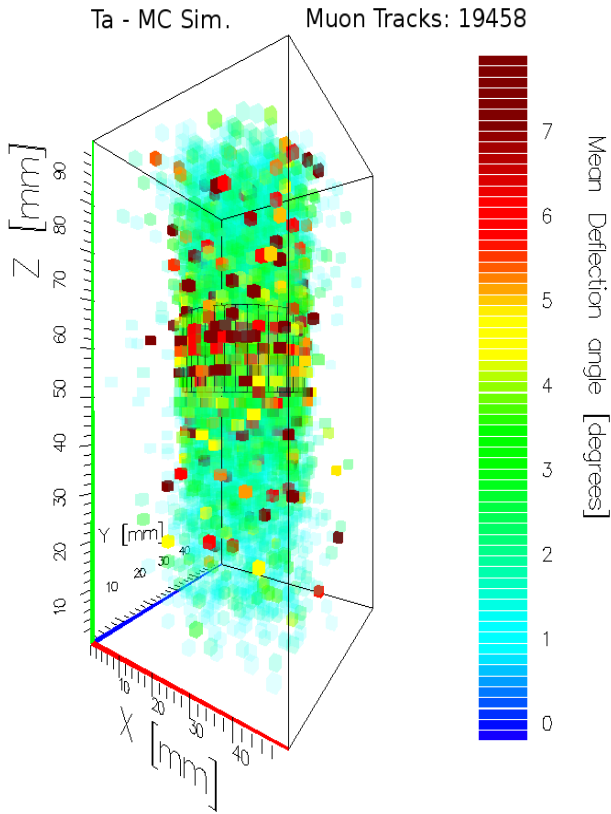

Fig. 5. Simulated MC tomogram of Ta cylinder with Vx4D using 19458 muon tracks. A cut of $\Theta_{mean} > 1°$ on the mean angle is applied for clarity.

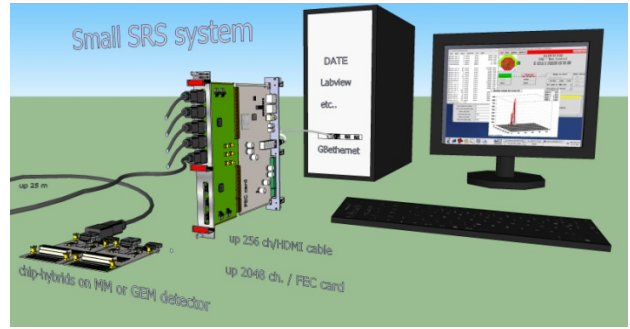

Fig. 6. Components of small Scalable Readout System (max. 16k channels).

### A. Frontend chip hybrid

The charge-sensitive amplification and readout chip is wire-bonded onto a small hybrid plug-in card (Fig. 7) that resides on the GEM detector. The first chip hybrids produced by us each carry a 128-channel, analog APV25 chip [21] and connect via standard 130-pin connectors to the GEM chamber. The hybrid board regulates power supplied via the HDMI cable and includes a PLL chip for clock refreshing and trigger decoding. Each channel has diode spark protection circuitry on the hybrid. The analog output signals of the APV chip are transmitted via a chip link to an ADC adapter card up to 25 m away. A single APV hybrid can be extended to a dual hybrid with 256 channels by linking it with a short cable to a neighboring "slave" hybrid card.

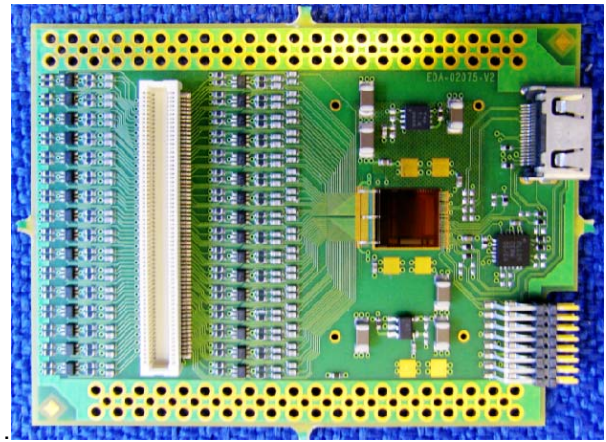

Fig. 7. Photo of the first SRS frontend hybrid with a bonded APV chip.

### B. ADC adapter for analog chips

SRS adapter cards interface the frontend chip via chip link cables to the SRS readout electronics. The first chip adapter produced by us is the CERN ADC adapter (Fig. 8) for analog frontend chips, such as the APV25 and Beetle 1.4 chips. Up to eight HDMI input connectors allow for up to 16 APV hybrids, i.e. 2048 channels, to be digitized by the ADC adapter card. The card implements two 12-bit octal-ADC chips with up to 50 (65) MS/s sampling rate and programmable gains. Different input cables lengths can be compensated via de-emphasis pre-stages.

*C. Readout with Frontend Concentrator card*

The Frontend Concentrator (FEC) card [22] was conceived as an RD51 partnership project between CERN and UP Valencia. Designed and produced by UPV, it is an FPGA-based system with SFP ethernet port and a NIM and/or LVDS trigger interface. An ADC card and an FEC card are electrically and mechanically interconnected via PCIe connectors that provide links with Gb speed, programmable I/O, I2C controls, and power. Up to 14 ADC/FEC pairs fit into one 6U×220 Eurochassis with each FEC card straddle-mounted to the adapter card in the same slot. As shown in Fig. 9, a mixture of A, B (3U) or C type (6U) adapter cards can be connected to the 6U FEC cards.

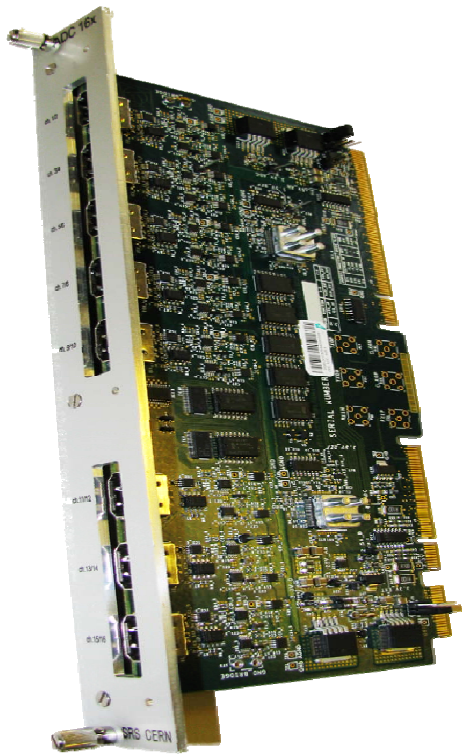

Fig. 8. Photo of the 6U SRS ADC adapter card for analog frontend chips.

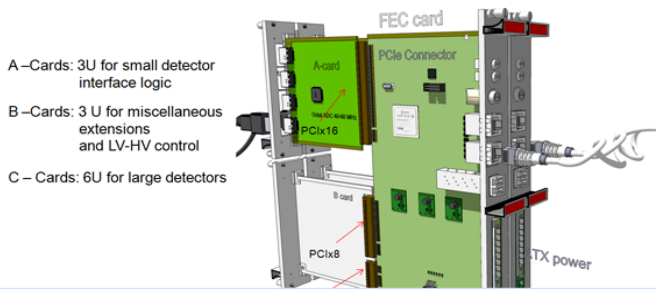

Fig. 9. Concept of SRS electronics cards with 3U or 6U adapter cards (A, B, C types) straddle-mounted via PCIe connectors to a 6U FEC card.

*D. Firmware and online monitoring*

The SRS firmware for the FEC card FPGA (Xilinx V5) consists of an application layer, which is specific to the ADC card and chip hybrid, plus a more universal DAQ interface layer with slow controls and ethernet core (Fig. 10). Data readout and slow controls are implemented over a UDP protocol between the FEC card and the online DAQ computer. For online monitoring we can use "Chipscope" [23] as provided by the Xilinx ISE suite. The CERN SRS team is also developing online monitoring via National Instrument's Labview (Fig. 11).

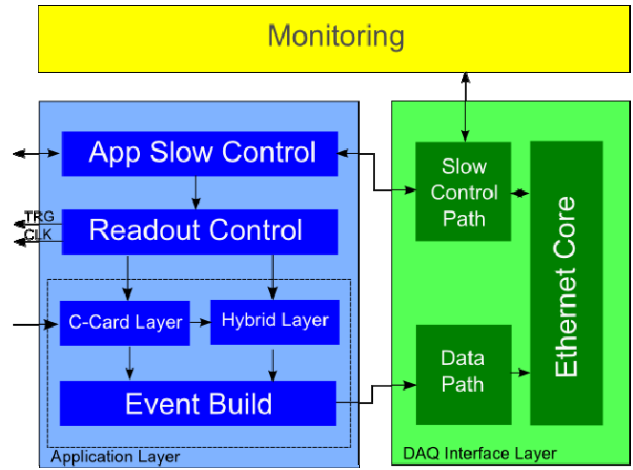

Fig. 10. Overview of SRS firmware.

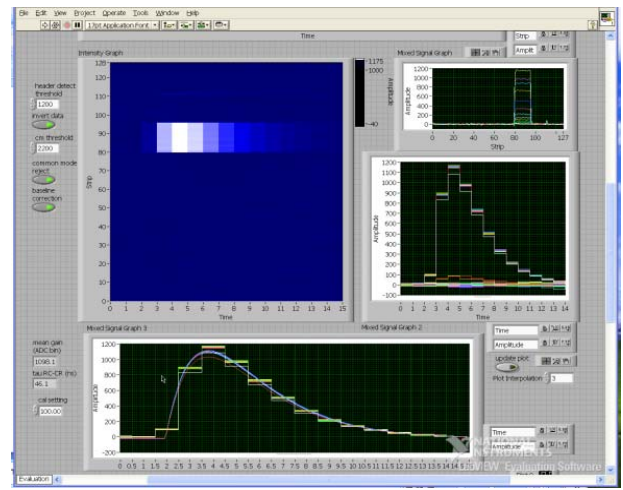

Fig. 11. Online DAQ monitoring with Labview.

*E. Data acquisition*

The UDP equipment port is an extension to the DATE online software of the ALICE experiment at the CERN Large Hadron Collider. It was developed by the ALICE DAQ group in parallel with the SRS developments and allows for several Gb ethernet links from individual FEC cards to be connected to a single online DAQ computer. Fig. 12 shows throughput and trigger rate as a function of the event payload size. With four connected streams of 1 Gbit/s, a throughput of 450 MByte/s is reached for event payloads above 10 kByte.

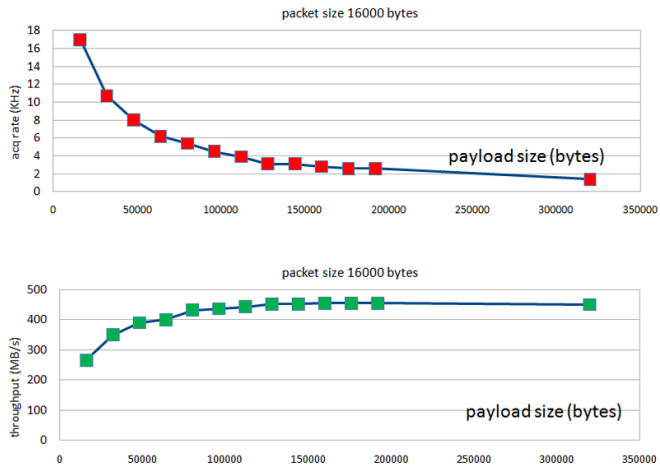

Fig. 12. Achievable SRS trigger rate vs. payload size (top). Bandwidth vs. payload size (bottom). Both quantities measured with 4 Gb data generators equivalent to four FEC cards connected to one online DAQ computer.

## IV. TRIPLE-GEM READOUT WITH SRS

### A. Experimental setup

The readout of entire $30 \times 30$ cm$^2$ Triple-GEM detectors for the prototype MT station constitutes the first field test of the full SRS readout chain from frontend hybrid and digitizer/FEC pair (Fig. 13) via Gb ethernet link over to online data acquisition and offline data analysis. In this first experimental setup, we read out a total of 1,536 microstrips (768 $x$-strips and 768 $y$-strips with 400 µm pitch) with 12 hybrid cards (Fig. 14). The SRS DAQ is triggered at a rate of 10 Hz by the coincidence signal produced by cosmic ray muons in a pair of $40 \times 50$ cm$^2$ plastic scintillators placed directly above and below the Triple-GEM. Within one day the whole chain of the SRS system can be made functional, enabling us to take cosmic data continuously for several days.

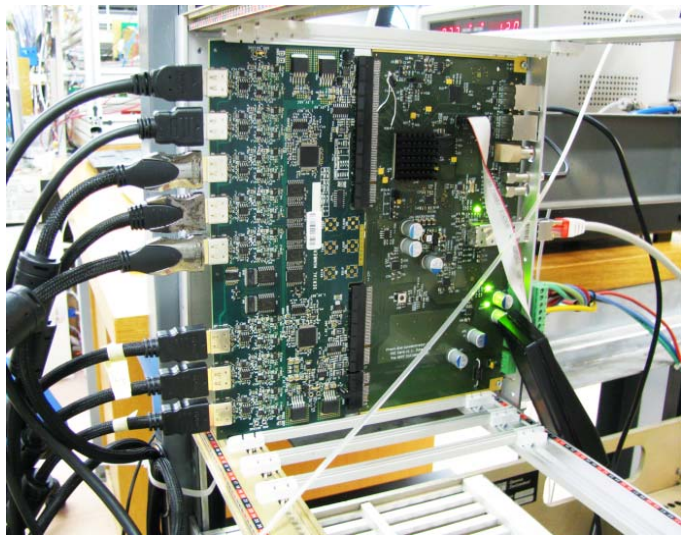

Fig. 13. Photo of powered and fully connected SRS ADC adapter card (left) and FEC card (right) in a 6U chassis.

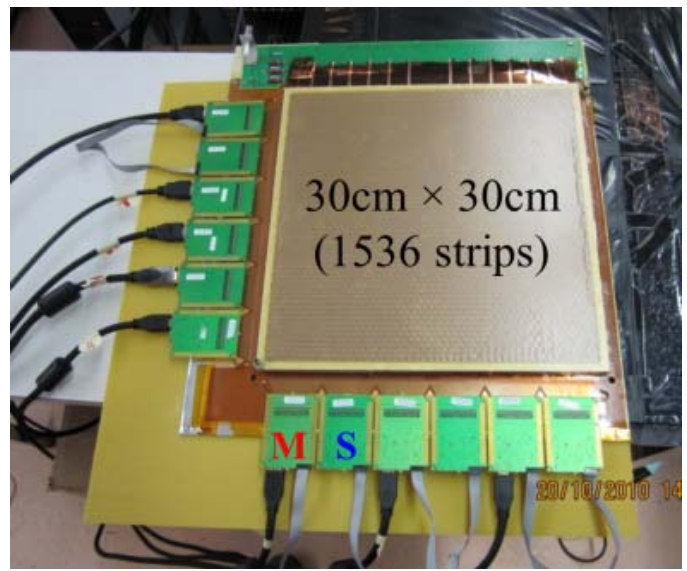

Fig. 14. Triple-GEM detector read out by 12 APV25 hybrids. A master (M) and a slave (S) APV hybrid card are indicated.

### B. First results from the SRS data for Triple-GEMs

We have recorded pedestal and cosmic data runs for five individual Triple-GEM detectors with the SRS DAQ for a total of 20k events in each run and an event size of 60 kByte without zero-suppression. The signals from the APV25 hybrids are sampled at a frequency of 40 MHz, which allows the hybrid to store up to 30 sampled time slices when operating in multi-peak mode. Fig. 15 shows typical APV25 raw event data with 14 consecutive 25 ns time frames of multiplexed digital data from 128 channels. Fig. 16 shows the time evolution of a cosmic ray muon signal in 25 ns time slices on the 128 strips read out by one APV25 after the raw data are decoded and strip numbers are properly mapped.

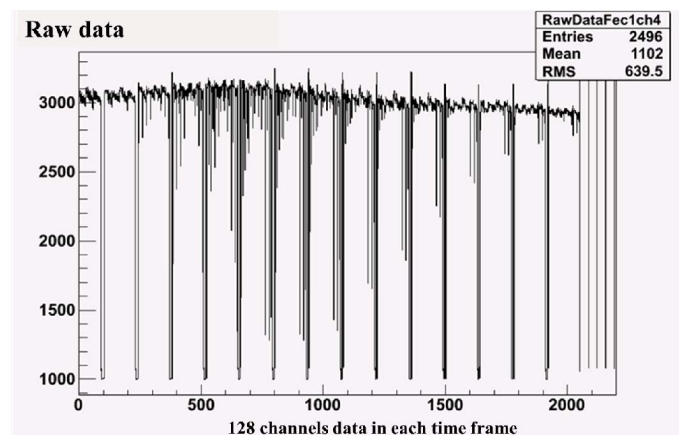

Fig. 15. Raw event data for digitized pulse heights in ADC counts vs. readout time frames as recorded with an APV25 hybrid for a cosmic ray muon signal.

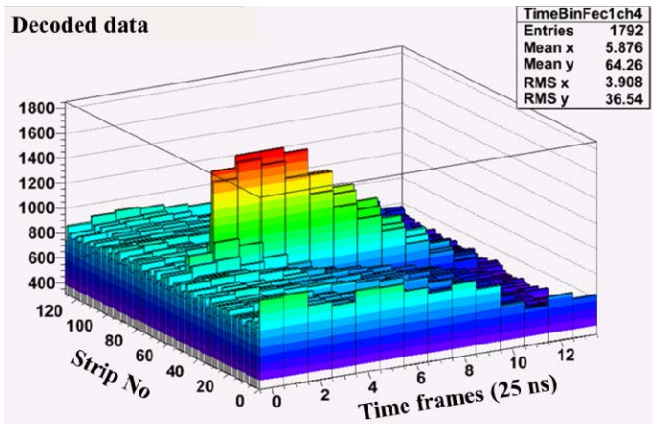

Fig. 16. Decoded event data from one APV25 hybrid for a cosmic ray muon.

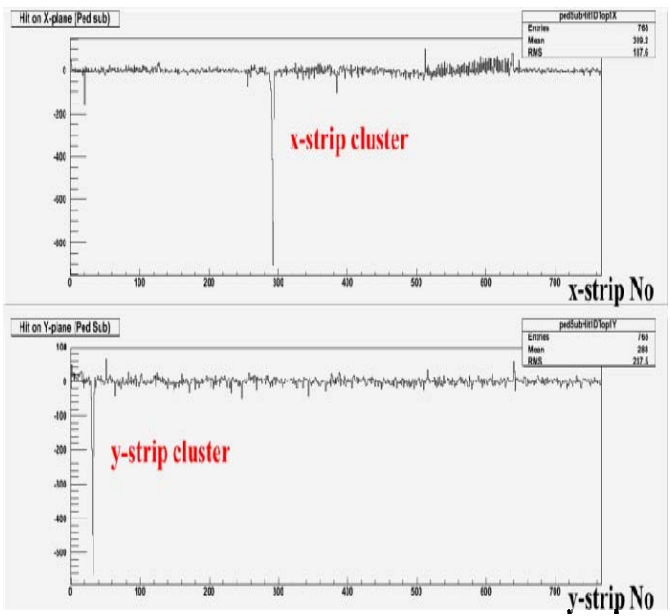

Fig. 17. Measured pulse height in ADC counts vs. strip number for *x*-strips (top) and *y*-strips (bottom) for a typical 2D hit cluster recorded with a Triple-GEM detector after baseline correction and pedestal subtraction are applied. A total of 1536 strips are read out with 12 APV hybrids.

We apply a baseline correction to the data for each event and for each time bin before subtracting the pedestals. From the observed pedestal widths, the average noise level is estimated to be ~20 ADC counts with a full ADC range of 4096 counts (12-bit ADC). The resulting muon events are found to have an average strip cluster size of about seven strips as shown in Fig. 17. The full set of cosmic ray muon data recorded for five different Triple-GEMs of the type shown in Fig. 14 is currently being analyzed.

## V. CONCLUSIONS

We have successfully produced 3D tomographic images of high-Z materials with the minimal MTS and conclude that GEM-based muon tomography is basically working. The performance of the SRS in its first field test of the full chain from GEM detector to offline analysis is very promising. Consequently, we are going ahead with production of the full SRS electronics required to equip a cubic-foot (~27 lit.) muon tomography station with ten Triple-GEM detectors, which will require a total of ~16k readout channels.

### ACKNOWLEDGMENT AND DISCLAIMER

We thank Leszek Ropelewski and the Gas Detector Development group, Pierre Vande Vyvre, Barthelemy von Haller, and Adriana Telesca from the ALICE DAQ group, all at CERN, for their help and technical support. We are also grateful to Maxim Titov and Esther Ferrer Ribas (CEA Saclay) for their help with the Gassiplex electronics for the minimal MTS.

This material is based upon work supported in part by the U.S. Department of Homeland Security under Grant Award Number 2007-DN-077-ER0006-02. The views and conclusions contained in this document are those of the authors and should not be interpreted as necessarily representing the official policies, either expressed or implied, of the U.S. Department of Homeland Security.

*A, 2010 Proc. 12th Symposium On Radiation Measurements and Applications*, Ann Arbor, MI, USA, arXiv:1007.0256 [physics.ins-det].

[15] S. Agostinelli, *et al.*, "GEANT4 – A simulation tool kit," *Nucl. Instrum. Meth. A,* vol. 506, pp. 250–303, 2003.

[16] http://root.cern.ch/drupal.

[17] http://root.cern.ch/root/html/tutorials/gl/glh3c.C.html.

[18] H. Muller, S. Martoiu, F. Costa, A. Tarazona, J. Toledo, and F. Zang, "The SRS scalable readout system for micro pattern gas detectors and other applications," article in preparation for submission for publication in *Nucl. Instr. and Meth. A,* 2010.

[19] V. Altini, *et al.,* "Commissioning and initial experience with the ALICE on-line," *2009 Proc. 17th International Conference on Computing in High Energy and Nuclear Physics*, J. Phys. Conf. Ser. 219:022022, 2010, Prague, Czech Republic.

[20] B. v. Haller, *et al.,* "The ALICE data quality monitoring," *2009 Proc. 17th International Conference on Computing in High Energy and Nuclear Physics*, J. Phys. Conf. Ser. 219:022023, 2010, Prague, Czech Republic.

[21] L. Jones, *et al.*, "The APV25 deep submicron readout chip for CMS detectors," *1999 Proc. 5th Conference on Electronics for LHC Experiments,* pp.162-166, Snowmass, CO, USA.

[22] J. Toledo, U. Politécnica Valencia, Spain, *private communication.*

[23] http://www.xilinx.com/tools/logic.htm.